# Scenario adaptive disruption prediction study for next generation burning-plasma tokamaks

J. Zhu[1], C. Rea[1], R.S. Granetz[1], E. S. Marmar[1], K. J. Montes[1], R. Sweeney[1], R.A. Tinguely[1], D. L. Chen[2], B. Shen[2], B. J. Xiao[2], D. Humphreys[3], J. Barr[3], O. Meneghini[3]

[1] Plasma Science and Fusion Center, Massachusetts Institute of Technology, Cambridge, MA, USA
[2] Institute of Plasma Physics, Chinese Academy of Sciences, Hefei, Anhui, China
[3] General Atomics, San Diego, CA, USA

E-mail: jxzhu@mit.edu

**Abstract**

Next generation high performance (HP) tokamaks risk damage from unmitigated disruptions at high current and power. Achieving reliable disruption prediction for a device's HP operation based on its low performance (LP) data is key to success. In this letter, through explorative data analysis and dedicated numerical experiments on multiple existing tokamaks, we demonstrate how the operational regimes of tokamaks can affect the power of a trained disruption predictor. First, our results suggest data-driven disruption predictors trained on abundant LP discharges work poorly on the HP regime of the same tokamak, which is a consequence of the distinct distributions of the tightly correlated signals related to disruptions in these two regimes. Second, we find that matching operational parameters among tokamaks strongly improves cross-machine accuracy which implies our model learns from the underlying scalings of dimensionless physics parameters like $q_{95}$, $\beta_p$ and confirms the importance of these parameters in disruption physics and cross machine domain matching from the data-driven perspective. Finally, our results show how in the absence of HP data from the target devices, the best predictivity of the HP regime for the target machine can be achieved by combining LP data from the target with HP data from other machines. These results provide a possible disruption predictor development strategy for next generation tokamaks, such as ITER and SPARC, and highlight the importance of developing on existing machines baseline scenario discharges of future tokamaks to collect more relevant disruptive data.

## 1. Introduction

Magnetic confinement of plasmas in tokamaks provides an attractive solution for controlled fusion energy production. To achieve steady state plasma confinement on next generation tokamaks, we need to reliably avoid or mitigate unexpected plasma terminations, called disruptions. Given the limited capability to predict disruptions [1] and availability of extensive experimental data from decades of tokamak operation, data-driven methods are strong candidates for disruption predictors. Various data-driven disruption predictors with good accuracy have been developed on JET [2-3], ASDEX-U [4-5], DIII-D [6-7], C-Mod [8] and JT-60U [9]. However, due to the large gap of dimensional and operational parameters between existing devices and next generation tokamaks, extrapolation of these predictors to near-future burning-plasma tokamaks, like ITER [10] and SPARC [11], is uncertain. So far, significant effort has been devoted to solving this problem. First, recent deep-learning based predictors have shown strong cross-machine ability [12-13] to learn general representations across tokamaks. Second, several previous studies [14-15] explored the strategy of building a predictor from scratch. In these studies, researchers gradually add data in chronological order to retrain the predictors, and then test on future unseen discharges. However, these studies are conducted using discharges from similar operational regimes which implicitly assumes that we can explore and learn on data that have similar parameters to future "test" discharges. Although this assumption is generally valid for existing machines, the ITER research plan [16] suggests this is probably not the case for future devices like ITER because unmitigated disruptions in High Performance (HP) regimes threaten the integrity of the facility, and we have to predict these disruptions using Low Performance (LP) data from these devices. However, data-driven predictors trained on data from initial LP campaigns might not work well on subsequent HP campaigns, due to the shift of plasma parameters, and thus the predictors trained on LP discharges are ineffective for HP discharges.

In this letter, by conducting numerical studies using data from existing tokamaks, we demonstrate three conclusions. First, explorative data analysis finds that constraining the ranges of a few parameters gives clear separation between the resulting LP and HP plasmas. The clear separation between LP and HP plasmas suggests LP regime physics is not sufficient for predicting HP regime disruptions and further implies the validity of the "*train-on-LP-data*" strategy [14-15] is unclear. Indeed, our results show data-driven predictors trained on abundant LP discharges work poorly on the HP regime of the same tokamak, which is a consequence of the distinct distributions of the tightly correlated signals related to disruptions in these two regimes. Second, our cross-machine studies show that matching operational parameters of other existing machines to the target device can greatly improve predictive accuracy for the target device, and progressively matching more parameters continually increases the predictive power for the target device. This conclusion implies our model learns from the underlying scalings of dimensionless physics parameters like $q_{95}$, $\beta_p$ and confirms the importance of these parameters in disruption physics and cross machine domain matching from the data-driven perspective. Finally, our results indicate that in the absence of HP data from the target devices, the best prediction results on the HP regime of the target device can be achieved by combining LP data from the target with HP data from other machines and training the predictor with this combined dataset. These conclusions provide the basis for developing a disruption prediction strategy for ITER and SPARC and highlight the importance of developing baseline scenario discharges on current devices and collecting more relevant data from these discharges to optimize the reliability of a disruption predictor for future HP devices. This letter focuses on ITER, but the same strategy would also be applicable to SPARC and other future HP devices.

## 2. *Using data from existing machines to simulate the LP and HP phases on ITER*

In the past few years, our group has created large disruption warning datasets on C-Mod, DIII-D and EAST [13, 17], and the same databases have been used for a previous cross-machine study [13]. Using these databases, our goal is to find one or more disruption prediction strategies that can be extrapolated to ITER.

At present, ITER's research plan incorporates a staged approach strategy aiming to increase the experimental capabilities that allow the transition to HP fusion operation [16, 18]. Following the completion of its first plasma, ITER will increase the toroidal magnetic field ($B_{tor}$), plasma current ($I_p$), density, and input power ($P_{in}$) toward final high $I_p$, $B_{tor}$ and $P_{in}$ fusion operation. Ramping up these parameters during ITER's early operation can pose challenges to the development of disruption predictors. First, the distributions of operational parameters - such as normalized plasma pressure ($\beta_p$), safety factor at the 95% flux surface ($q_{95}$), and Greenwald density ($n_G$) - will change with the increasing $I_p$, $B_{tor}$ and $P_{in}$ and thus might affect the power of the trained predictor. Second, due to the potentially serious damage to the device from high-current, high-stored-energy disruptions, the ITER research plan requires developing a reliable Disruption Mitigation System (DMS) trigger before the beginning of HP operation [16]. This requirement can result in differences between the training and testing operational regimes of the DMS trigger and may invalidate the disruption prediction algorithm.

To simulate the discrepancy between the training and testing domains of the ITER DMS trigger, among all changing operational parameters, we select three parameters: $\beta_p$, $P_{in}$ (not a training feature) and $q_{95}$ that are closely related to tokamak operation but less significant to disruption prediction on three tokamaks we studied [17] and calculate their $I_p$-flattop-averaged values[1] for each plasma discharge in our databases. From the distributions of flattop averaged parameters, we choose low/high cutoff thresholds for all three parameters (table 1) and select various **LP/HP (low/high $\beta_p$, low/high $P_{in}$, high/low $q_{95}$) datasets** on three devices. The chosen cutoff thresholds vary for different devices and depend on the distributions of each signal on the different devices as well as on suggestions from tokamak operators. Notice that the three chosen parameters are a small subset of all signals used for prediction models [13]; to see how limiting the ranges of three chosen parameters affect distributions of other training features, an orthogonal linear transformation called Principal Component Analysis (PCA) [19] is applied to all 12 training features (including $q_{95}$, $\beta_p$) of the combined LP and HP datasets for all three devices. Before applying the PCA transformation, each signal in this combined dataset is separately normalized to mean 0 and standard deviation 1 such that two principal components are not dominated by $q_{95}$ and $\beta_p$. In figure 1, each magenta point represents a 10 time-step sequence of 12 training features randomly sampled from the flattop of a HP shot while each cyan point represents a sequence randomly sampled from the flattop of a LP shot. The two principal components (x, y axes) are linear combinations of 12 training features and our PCA analysis suggests that the 10 unconstrained features have significant contribution to them. Thus, if the joint distribution of unconstrained features is not strongly affected by three chosen parameters (similar for LP and HP plasmas), the distributions of resulting LP and HP plasmas in the projected 2-D plane should have a large overlap. However, the PCA clustering plots show that there is only tiny overlap between the resulting LP/HP plasmas for all three devices, which implies signals related to disruption prediction are closely correlated. Limiting the ranges of a few less significant parameters can greatly change distributions of other signals related to disruption prediction and makes clear distinction between LP and HP plasmas. This observation further implies LP regime physics is too limited and does not have enough overlap with the HP regime physics to adequately train the predictor. More PCA plots for different subdivisions of HP data can be found in appendix which further support our conclusion about the signal correlation.

---
[1] For $\beta_p$ and $P_{in}$, the average is only computed during the flattop $I_p$ period when external heating is active.

**Table 1: Performance cutoff threshold of $\beta_p$, $P_{in}$ and $q_{95}$ on three devices**

|        | $\beta_p$ low/high cutoff | | $P_{in}$ low/high cutoff (MW) | | $q_{95}$ low/high cutoff | |
|--------|------|------|------|------|------|------|
| C-Mod  | <0.15 | >0.25 | <1.0 | >3.0 | <4.0 | >4.6 |
| DIII-D | <0.60 | >0.80 | <3.5 | >7.5 | <4.5 | >5.0 |
| EAST   | <0.55 | >0.75 | <0.6 | >3.0 | <5.0 | >6.0 |

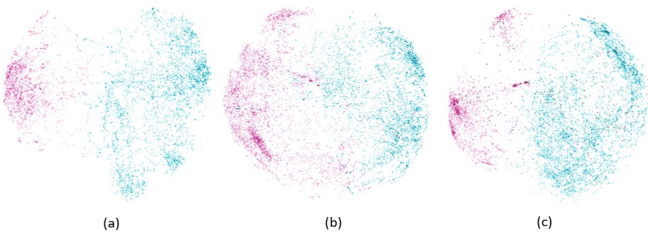

(a)        (b)        (c)

**Figure 1: The PCA clustering plots for: (a) C-Mod; (b) DIII-D; and (c) EAST. The coloring is done *a posteriori*.**

Another component of our study is the disruption predictor. In our previous research, we have developed a Hybrid Deep-Learning (HDL) disruption predictor that achieves state-of-the-art accuracy on multiple tokamaks with only limited hyperparameter tuning [13]. The HDL framework consists of two parts: (i) A deep neural-net that converts an input plasma sequence to an output 'disruptivity' label between 0 (non-disruptive class) and 1 (disruptive class); (ii) a testing scheme to trigger an alarm on a shot-by-shot basis. By changing the disruptivity threshold of the trained HDL predictor, we can make a trade-off between the fraction of false alarms (False Positive Rate, FPR) and the fraction of successfully detected disruptions (True Positive Rate, TPR). This trade-off is characterized by the receiver operator characteristic (ROC) curve [16]. Throughout this letter, we will use the HDL predictor to conduct all numerical disruption studies, and the result of each experiment is evaluated using the ROC curves at 50ms before the final current quench. However, since all data-driven methods essentially learn from the empirical distribution of the input signals, we argue that our analysis is applicable and general to all data-driven methods. 50ms is chosen since it is the warning time required for the ITER DMS [20].

## 3. *Scenario based cross-machine study*

Given the risks for significant damage to the devices from unmitigated high-current, high-stored-energy disruptions, the development of a disruption predictor for HP burning-plasma operations, using only LP data from the device, is one of the suggested approaches for ITER. Based on this approach, a "*train-on-LP-data*" strategy based on literature [14-15] is to train a predictor using data from the early stages of ITER's operation and apply it to subsequent discharges. If the predictor trained on initial LP ITER data can learn knowledge that is applicable to HP regime of the device, it should be able to predict disruptions in HP regime. Using the HDL predictor and various LP/HP datasets from three devices, we can investigate whether this "*train-on-LP-data*" strategy works. If not, given the strong cross-machine potential of deep-learning-based predictors, we seek to improve target prediction accuracy by using data from other devices. Here, we consider C-Mod and EAST as '*existing/other machines*', with DIII-D chosen as the '*new/target device*' and conduct numerical experiments to explore the best strategy of developing data-driven predictors that can predict disruptions on the HP regime of the *new device* using only LP data from the *new device,* combined with HP data from the *existing machines*. The training and testing set compositions of all experiments can be found in table 2. In addition, **all following qualitative conclusions are machine-independent**: they always hold no matter which device is selected as the '*new* *device'*. The other two permutations are shown in the appendix.

**Table 2. Training and testing set composition of all experiments using DIII-D as the '*new machine*'**

| Case No. | Training set | Testing set |
|---|---|---|
| 1 | 209 DIII-D LP ($\beta_p$<0.6, $P_{in}$<3.5MW, $q_{95}$>5) shots (14% disruptive) | 240 DIII-D HP ($\beta_p$>0.8, $P_{in}$>7.5MW, $q_{95}$<4.5) shots (15% disruptive) |
| 2 | 209 DIII-D high $q_{95}$ (>5) shots (5% disruptive) | |
| 3 | 100 DIII-D HP shots (15% disruptive) | 140 DIII-D HP shots[2] (15% disruptive) |
| 4 | 100 DIII-D HP low $B_{tor}$ (<1.79T) shots (20% disruptive) | 140 DIII-D HP, high $B_{tor}$ (>1.79T) shots[3] (12% disruptive) |
| 5 | 100 DIII-D high $\beta_p$ (>0.8), low $q_{95}$ (<4.5), low $B_{tor}$ (<1.79T) shots (14% disruptive) | |
| 6 | 209 C-Mod low $q_{95}$ (<4) shots | Same as cases 1-2 |
| 7 | 209 C-Mod high $q_{95}$ (>5) shots | |
| 8 | 209 EAST low $q_{95}$ (<5) shots | |
| 9 | 209 EAST high $q_{95}$ (>6) shots | |
| 10 | 209 EAST low $q_{95}$ (<5) high $\beta_p$ (>0.4) shots | |
| 11 | 209 C-Mod low $q_{95}$ (<4) shots plus 209 DIII-D LP shots | |
| 12 | 209 EAST low $q_{95}$ (<5) high $\beta_p$ (>0.4) shots plus 209 DIII-D LP shots | |
| 13 | 209 C-Mod high $q_{95}$ (>5) shots plus 209 DIII-D LP shots | |
| 14 | 209 EAST high $q_{95}$ (>6) shots plus 209 DIII-D LP shots | |

The first set of numerical experiments is conducted using only data from our target *new device* to test the effectiveness of the "*train-on-LP-data*" strategy. The results of these experiments (cases 1-5) are shown in figure 2(a)-(b). The training and testing set composition of these cases can be found in table 2. From the results of these cases, it is possible to draw the following conclusions:

1. Limiting the ranges of chosen parameters in the training set strongly affects the test performance of trained data-driven predictor. A predictor trained on a few hundred high $q_{95}$ discharges works poorly for the HP regime of the same device (case 2 in figure 2(a)). Furthermore, as more parameters ($\beta_p$, $P_{in}$) of the training discharges deviate from the target HP regime, the predictor's accuracy for the HP regime becomes continuously worse (case 1 in figure 2(a)). Notice that there are different numbers of disruptive discharges in different training sets. Although having only a small number of disruptive samples in the training set can decrease the accuracy of trained predictor, our experiment results suggest this is just a secondary effect compared with the effect of operational regime. Despite having the most disruptive training shots (~30 disruptive training shots in case 1), case 1 gives the worst test accuracy among cases 1-3. Given all these results, we conclude that a predictor trained on abundant LP discharges performs poorly for the HP regime of the same device.

2. A data-driven predictor can effectively learn disruption physics if the training and test data come from similar operational regimes. A predictor trained on only 100 HP shots of the target device already achieves the best test accuracy among cases 1-3 (case 3 in Fig. 2(a)-(b)).

From the first conclusion above, even without other constraints to the training set, the $q_{95}$ discrepancy between the training and testing sets can significantly decrease the prediction accuracy for the target HP regime. Since high current disruptions can be dangerous to ITER, we need to develop a predictor using only low current ITER discharges. Under this constraint, to match $q_{95}$ between training and testing regime, a possible way is to train a predictor on low $B_{tor}$, low current and low $q_{95}$ discharges. To test this, we sub-select low $B_{tor}$ shots from the *new device* HP database as the training set, and test on the remaining HP high $B_{tor}$ *new device* data. This is the fourth case. Moreover, since selecting high $P_{in}$ shots from low $B_{tor}$ discharges can give highly skewed dataset, in the fifth case, the predictor is trained on low $q_{95}$, high $\beta_p$ and low $B_{tor}$ *new device* shots and tested on HP high $B_{tor}$ *new device* data. In figure 2(b), we compare the results of cases 4-5 with case 3 (training and testing data from the same HP regime) which gives the following additional conclusions:

3. Although $P_{in}$ and $B_{tor}$ are not training features, predictors trained on HP (with/out $P_{in}$ constraint) low $B_{tor}$ discharges perform poorly for the HP high $B_{tor}$ discharges. This implies the ranges of parameters like $B_{tor}$ and $P_{in}$ can greatly affect the feature space of predictors even if they are **not** training features, and ITER would need to reach relatively **high** $B_{tor}$ as early as possible during its LP pre-fusion phase, even with low current and high $q_{95}$ (cases 3-5 in figure 2(b)).

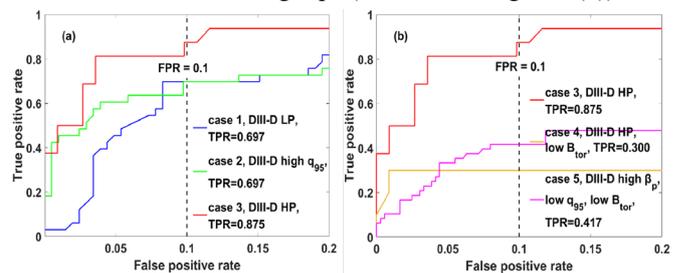

**Figure 2: ROC curves from the *new device* (DIII-D) test set using only *new device* data. The training and testing set compositions of all cases can be found in table 2.**

From the existing literature [17] and our previous HDL studies, $q_{95}$ and $\beta_p$ are not the signals with the most significance to disruption prediction on the three tokamaks we studied. Therefore, directly learning from these constrained features is not required for achieving high prediction rate and the discrepancies of $P_{in}$, $q_{95}$ and $\beta_p$ ranges between LP and HP data themselves will not lead to significantly worse prediction accuracy on the test set. Since most of the training features (10/12) are not artificially constrained, if limiting the ranges of three chosen parameters does not significantly change distributions of other parameters, the predictor trained on the resulting LP data (case 1) should work well on HP test set (close to the result of case 3). However, the above results show predictor trained on LP data works significantly worse on HP test set. This observation again suggests signals related to disruption prediction are strongly correlationed. Although the chosen physics-based signals ($\beta_p$, $P_{in}$, $q_{95}$) do not directly contribute much to the power of model, limiting their range can deeply affect the distributions of more important signals and hence change the prediction results. Thus, without additional data, developing a disruption predictor that works for the HP regime of tokamak using only LP data from the same device is unlikely to work because LP regime physics is too limited to predict disruptions in HP regime.

---

[2]For case 3, 140 DIII-D HP shots was selected from the total 240 DIII-D HP shots as the test set and the remaining 100 DIII-D HP shots was used as the training set. 11 independent experiments were run for case 3 (11 different random partition of 240 DIII-D HP shots). The case 3 result shown in figure 2 correspond to the median accuracy among 11 results which makes it comparable with the results of other cases.
[3]For cases 4-5, 140 DIII-D HP high $B_{tor}$ shots was selected from 240 DIII-D HP shots as the test set to evaluate the effect of $B_{tor}$.

To seek a better strategy, we conducted another set of numerical experiments, using data from both the *new device* and *existing machines*. Given the results from the first set of numerical experiments, an attempt is made to match parameters between the *new device* and the *existing machines*. The results of these experiments (cases 6-12) are shown in figures 3(a)-(d). The training and testing set composition of these cases can be found in table 2. The results of figure 3 point to the following conclusions:

4. A predictor trained on low $q_{95}$ data from *existing machines* performs better for the HP regime of the *new device* (cases 6, 7 in figure 3(a), cases 8, 9 in figure 3(b)). Training predictor using low $q_{95}$ *and* relatively high $\beta_p$ data from *existing machines* further increases the accuracy on the HP regime of the *new device* (case 10 in figure 3(b)). These results demonstrate that training on "matched" data (with similar operational parameters to the test regime) from existing machines greatly outperforms the unmatched data, and progressively matching more operational parameters continuously improves the target performance. Therefore, developing ITER baseline scenario discharges on existing tokamaks, and training predictors on these, should greatly improve disruption prediction on ITER itself.

5. In the absence of HP data from the *new device*, combining "matched" HP data from *existing machines* with LP data from the *new device* gives the best prediction rate for the HP regime of the *new device* while adding "unmatched" data from *existing machines* to the training set can even decrease prediction accuracy on the *target device* (case 11, 13 in figure 3(c), case 12, 14 in figure 3(d)).

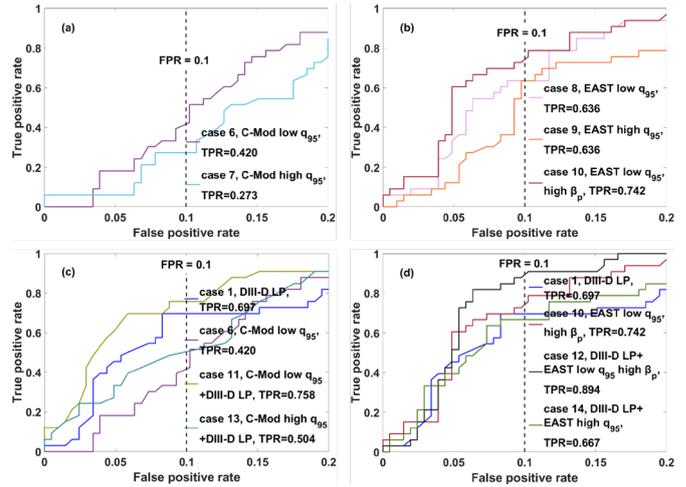

**Figure 3: ROC curves from the *new device* (DIII-D) test set using both *new device* and *existing machines* (C-Mod, EAST) data. The training and testing set compositions of all cases can be found in table 2.**

Given all conclusions from two sets of numerical experiments, we conclude that, due to the distinct distributions of the tightly correlated signals related to disruptions in HP and LP regimes, any data-driven predictors trained on early LP ITER data cannot be directly applied to future HP operation. Our analysis shows that developing a reliable DMS trigger for ITER's HP operation, using only LP data from ITER itself, requires HP ITER baseline discharges from existing machines. A possible strategy for ITER DMS trigger development is as follows: combine ITER LP data (low $\beta_p$, low $P_{in}$, high $q_{95}$ with relatively high $B_{tor}$) with HP ITER baseline discharges from other devices to train a predictor with enough accuracy to help ITER conserve its disruption budget during the early stage of its HP operation; as ITER's HP operation continues, add HP ITER data to the training set. Retraining the predictor using this combined dataset will boost the predictor performance towards ITER's long-term requirements [1].

## 4. *Summary and future plans*

Given the risks of significant damage to fusion devices from unmitigated high-current, high-power disruptions, developing a DMS trigger for HP burning-plasma operation before starting HP campaigns is crucial for the success of next generation tokamaks. In this letter, using databases from C-Mod, DIII-D and EAST, we selected three parameters that are closely related to tokamak operation but less significant to disruption prediction on the three tokamaks and built LP/HP datasets that can simulate the LP and HP phases on ITER. Our preliminary data exploration on these datasets finds limiting the ranges of three chosen parameters clearly separates the resulting LP/HP plasmas which indicates LP regime physics is insufficient for predicting HP regime disruptions. Dedicated numerical experiments based on these datasets further demonstrate that although a data-driven predictor can effectively learn when training and testing data come from the HP regime of the same device, having even one parameter of the training set deviate from the test operational regime greatly decreases the test performance of the trained predictors. Since $q_{95}$ and $\beta_p$ are not the top significant signals to the HDL model on the three machines we studied, the above results suggest different signals related to disruption prediction are strongly correlated. Therefore, pushing the limits of less important signals changes the distributions of more significant signals and thus decrease the power of trained predictor. Thus, any data-driven predictors trained only on LP discharges can perform poorly for the subsequent HP regime of the same tokamak, which suggests the "*train-on-LP-data*" strategy is not suitable for ITER.

Our cross-machine numerical experiments show that matching operational parameters among devices can greatly improve prediction accuracy for the target device. In the absence of HP data from the target devices, the best prediction results on the HP regime of the target device can be achieved by training the predictor on LP data from the target plus HP data from other machines. This conclusion implies that our model learns from the underlying scalings of dimensionless physics parameters like $q_{95}$, $\beta_p$ and confirms the importance of these parameters in disruption physics and cross machine domain matching from the data-driven perspective. Given all above findings, we conclude combining burning-plasma simulation discharges from experiments on existing tokamaks, with initial LP data from the next step device, is a promising strategy for the development of a DMS trigger for the next step tokamaks. Thus, the development of DMS trigger for future burning-plasma devices requires us to build comprehensive databases that consist of different kinds of disruptive burning-plasma baseline scenario discharges from current devices. Developing burning-plasma baseline scenarios on existing machines and exploring different kinds of disruptions that can happen during next step device's HP operation in the burning-plasma baseline scenarios of current devices to collect relevant data is crucial for improving disruption prediction on future devices. The next step of this study will be to extend our analysis to other existing tokamaks like JET, AUG and KSTAR. In particular, we will investigate whether our findings hold for the ITER baseline discharges on existing tokamaks.

## Acknowledgements

This material is based upon work supported by the U.S. Department of Energy, Office of Science, Office of Fusion Energy Sciences, using the DIII-D National Fusion Facility, a DOE Office of Science user facility, under Awards DE-FC02-04ER54698 and DE-SC0014264. Additionally, this work is supported by the National MCF Energy R&D Program of China, Grant No. 2018YFE0302100. The HDL architecture reported in the letter was developed using TensorFlow library [21]. The authors are grateful to N. Logan and C. T. Holcomb for their supports and valuable discussions.

## Disclaimer

Appendix:

1. Distributions of flattop averaged $\beta_p$, $P_{in}$ and $q_{95}$ on C-Mod, DIII-D and EAST

In figure 1-3, we show the distributions of flattop averaged $\beta_p$, $P_{in}$ and $q_{95}$ on C-Mod, DIII-D and EAST. In all subplots, the red lines give the high cutoff threshold of three signals while black lines give the low cutoff threshold of three signals. As can be seen from these plots, our cutoff thresholds well separate the low/high values of each signal and also leave enough data in the extracted low ends and high ends.

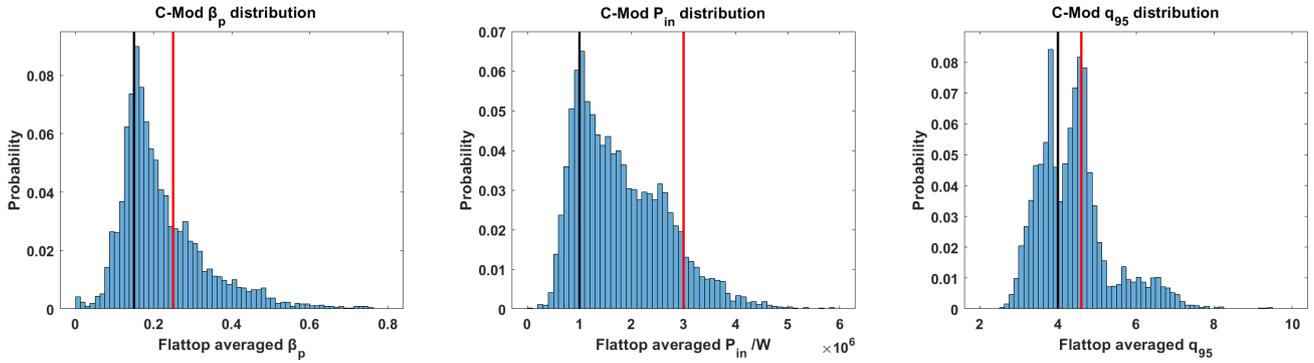

figure 1: Distributions of $\beta_p$, $P_{in}$ and $q_{95}$ on C-Mod. In all subplots, the red lines give the high cutoff threshold of each signal while black lines give the low cutoff threshold of each signal.

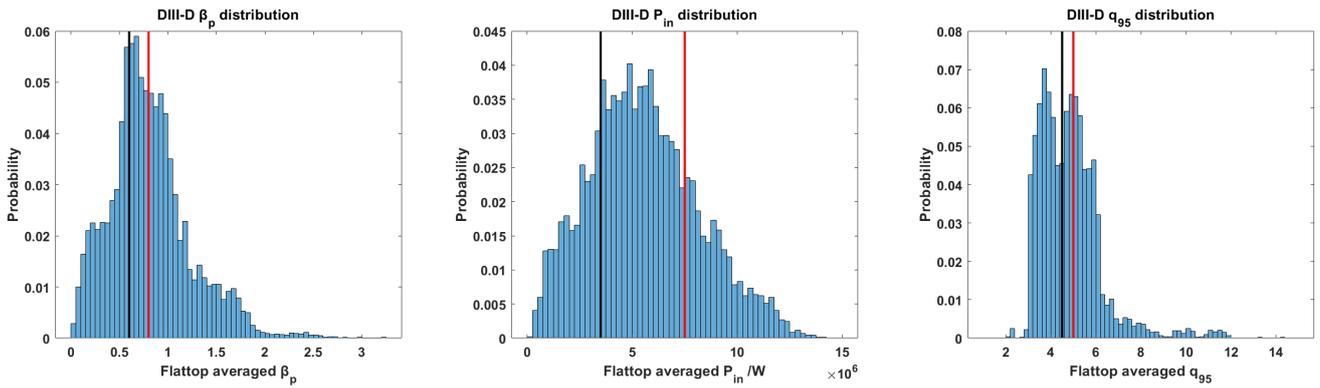

figure 2: Distributions of $\beta_p$, $P_{in}$ and $q_{95}$ on DIII-D. In all subplots, the red lines give the high cutoff threshold of each signal while black lines give the low cutoff threshold of each signal.

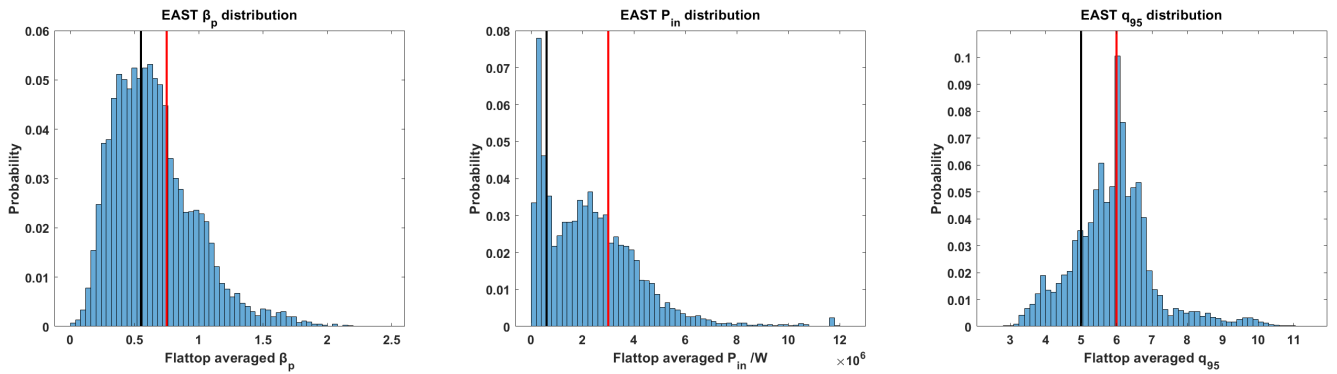

figure 3: Distributions of $\beta_p$, $P_{in}$ and $q_{95}$ on EAST. In all subplots, the red lines give the high cutoff threshold of each signal while black lines give the low cutoff threshold of each signal.

2. Numerical experiments using C-Mod as the '*new device*'

In these experiments, we consider DIII-D and EAST as '*existing machines*' with C-Mod chosen as '*new device*', the results are shown in appendix figure 4-5 which consistently support our previous cross-machine conclusions presented in Section 3, Scenario based cross-machine study. The training set compositions of these experiments are summarized in table 1. Notice that since C-Mod has much lower $\beta_p$ compared with DIII-D and EAST, to match $\beta_p$ range of C-Mod HP regime, we need to choose low $q_{95}$, low $\beta_p$ discharges from DIII-D and EAST. Furthermore, we find predictor trained on C-Mod's LP discharges cannot predict (the accuracy of trained predictor is close to random guessing) disruptions on C-Mod's HP regime which is consistent with our observation from the C-Mod PCA plot (figure 1(a)) because there is nearly no overlap between C-Mod LP (cyan) and HP (magenta) clusters.

**Table 1 training and testing set composition of all experiments using C-Mod as the '*new machine*'**

| Case No. | Training set | Testing set |
|---|---|---|
| 1 | 246 C-Mod LP ($\beta_p$<0.15, $P_{in}$<1MW, $q_{95}$>4.6) shots (12% disruptive) | 168 C-Mod HP ($\beta_p$>0.25, $P_{in}$>3MW, $q_{95}$<4) shots (19% disruptive) |
| 2 | 246 C-Mod high $q_{95}$ (>4.6) shots (12% disruptive) | |
| 3 | 75 C-Mod HP shots (19% disruptive) | 93 C-Mod HP (19% disruptive) |
| 4 | 75 C-Mod HP low $B_{tor}$ (<5.47T) shots (24% disruptive) | 93 C-Mod HP high $B_{tor}$ (>5.47T) shots (15% disruptive) |
| 5 | 75 C-Mod high $\beta_p$ (>0.25), low $q_{95}$ (<4), low $B_{tor}$ (<5.47T) shots (23% disruptive) | |
| 6 | 246 DIII-D high $q_{95}$ (>5) shots | Same as cases 1-2 |
| 7 | 246 DIII-D low $q_{95}$ (<4.5) shots | |
| 8 | 246 DIII-D low $q_{95}$ (<4.5) low $\beta_p$ (<0.6) shots | |
| 9 | 246 EAST high $q_{95}$ (>6) shots | |
| 10 | 246 EAST low $q_{95}$ (<4.5) shots | |
| 11 | 246 EAST low $q_{95}$ (<4.5), low $\beta_p$ (<0.5) shots | |
| 12 | 246 DIII-D low $q_{95}$ (<4.5), low $\beta_p$ (<0.6) shots plus 246 C-Mod LP shots | |
| 13 | 246 EAST low $q_{95}$ (<4.5), low $\beta_p$ (<0.5) shots plus 246 C-Mod LP shots | |
| 14 | 246 DIII-D high $q_{95}$ (>5) shots plus 246 C-Mod LP shots | |
| 15 | 246 EAST high $q_{95}$ (>6) shots plus 246 C-Mod LP shots | |

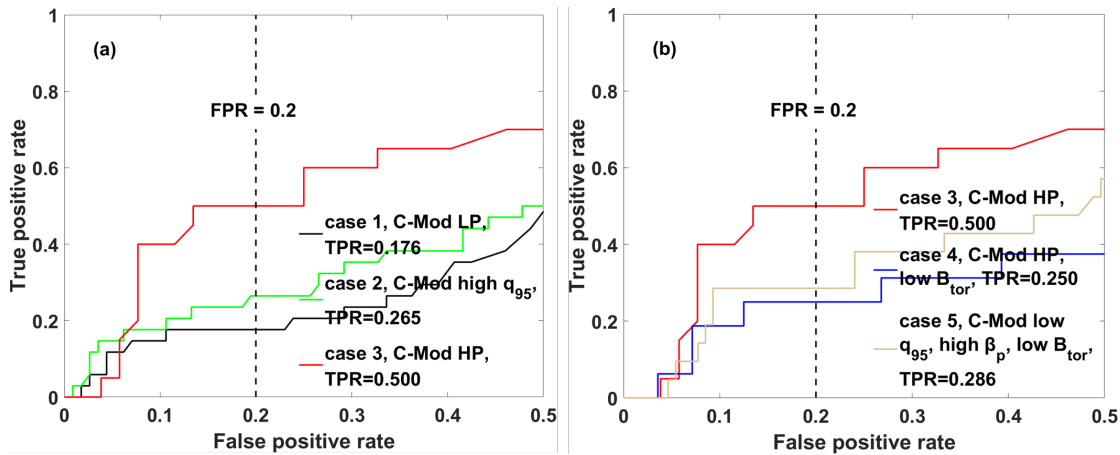

**Figure 4:** ROC curves from the *new device* (C-Mod) test set using only *new device* data. The training and testing set compositions of all cases can be found in table 1.

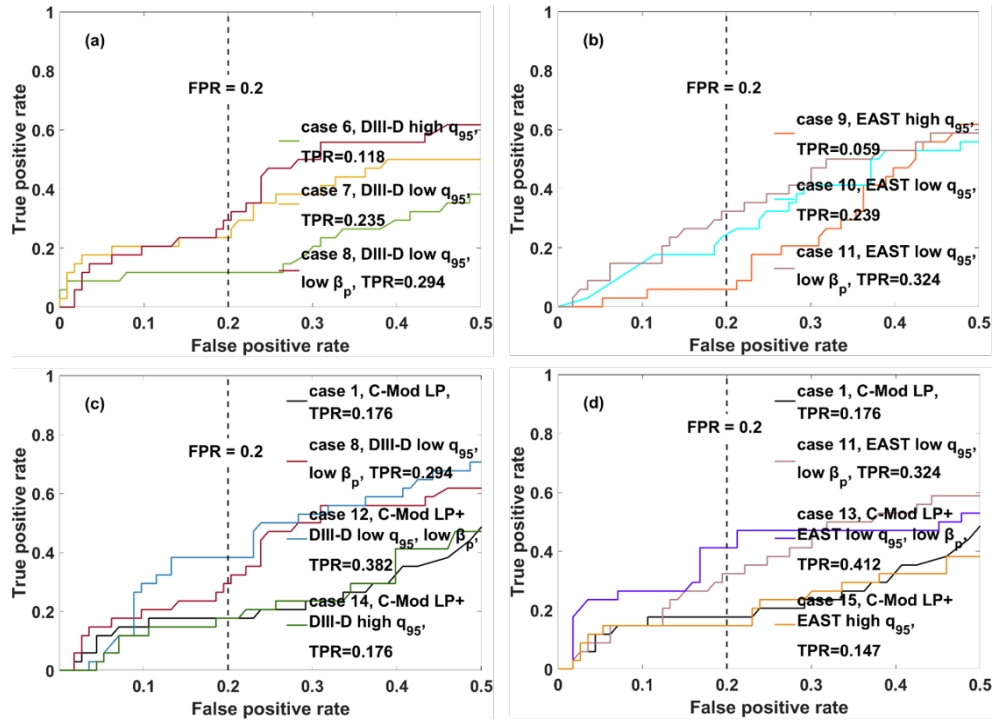

**Figure 5:** ROC curves from the *new device* (C-Mod) test set using both *new device* and *existing machines* (DIII-D, EAST) data. The training and testing set compositions of all cases can be found in table 1.

3. Numerical experiments using EAST as the '*new device*'

In these experiments, we consider C-Mod and DIII-D as 'existing machines' with EAST chosen as 'new device', the results are shown in appendix figure 6 and figure 7 which consistently support our previous cross-machine conclusions presented in Section 3, Scenario based cross-machine study. The training set compositions of these experiments are summarized in table 2.

**Table 2 training and testing set composition of all experiments using EAST as the '*new machine*'**

| Case No. | Training set | Testing set |
|---|---|---|
| 1 | 501 EAST LP ($\beta_p$<0.55, $P_{in}$<0.6MW, $q_{95}$>6) shots (43% disruptive) | 232 EAST HP ($\beta_p$>0.75, $P_{in}$>3MW, $q_{95}$<5) shots (47% disruptive) |
| 2 | 501 EAST high $q_{95}$ (>6) shots (28% disruptive) | |
| 3 | 100 EAST HP shots (47% disruptive) | 132 EAST HP shots (47% disruptive) |
| 4 | 100 EAST HP low $B_{tor}$ (<1.66T) shots (55% disruptive) | 132 EAST HP high $B_{tor}$ ($B_{tor}$>1.66T) shots (41% disruptive) |
| 5 | 100 EAST high $\beta_p$ (>0.75), low $q_{95}$ (<5), low $B_{tor}$ (<1.66T) shots (53% disruptive) | |
| 6 | 250 C-Mod high $q_{95}$ (>6) shots | Same as cases 1-2 |
| 7 | 250 C-Mod low $q_{95}$ (<5) shots | |
| 8 | 250 DIII-D high $q_{95}$ (>6) shots | |
| 9 | 250 DIII-D low $q_{95}$ (<5) shots | |
| 10 | 250 DIII-D low $q_{95}$ (<5) high $\beta_p$ (>0.4) shots | |
| 11 | 250 C-Mod low $q_{95}$ (<5) shots plus 501 EAST LP shots | |
| 12 | 250 DIII-D low $q_{95}$ (<5) high $\beta_p$ (>0.4) shots plus 501 EAST LP shots | |
| 13 | 250 C-Mod high $q_{95}$ (>6) shots plus 501 EAST LP shots | |
| 14 | 250 DIII-D high $q_{95}$ (>6) shots plus 501 EAST LP shots | |

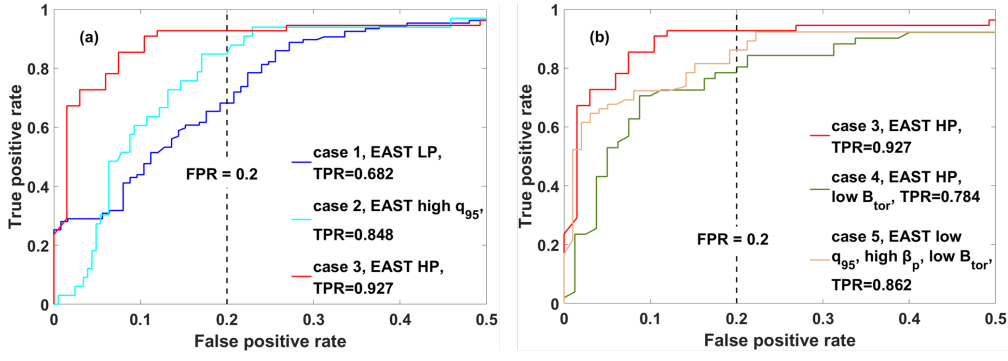

**Figure 6:** ROC curves from the *new device* (EAST) test set using only *new device* data. The training and testing set compositions of all cases can be found in table 2.

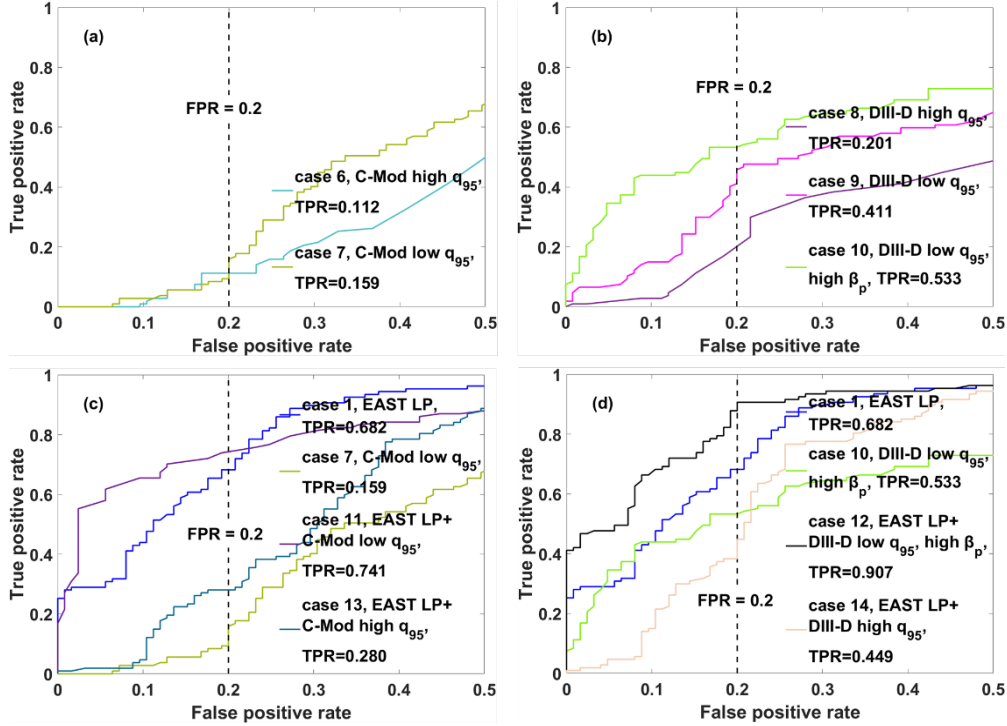

**Figure 7:** ROC curves from the *new device* (EAST) test set using both *new device* and *existing machines* (C-Mod, DIII-D) data. The training and testing set compositions of all cases can be found in table 2.

4. PCA analysis for different subdivisions of training sets in table 2 of the main text

In figures A8-A10, we present the PCA analysis of (a) HP low $B_{tor}$ data (i.e. case 4 training data, cyan) vs. HP high $B_{tor}$ data (i.e. case 4-5 test data, magenta) and (b) high $\beta_p$, low $q_{95}$, low $B_{tor}$ data (i.e. case 5 training data, cyan) vs. HP high $B_{tor}$ data (i.e. case 4-5 test data, magenta) for three machines we studied. Again, we want to point out the PCA transformation is applied to all 12 training features (includes $q_{95}$, $\beta_p$) of our model. Both $B_{tor}$ and $P_{in}$ are not one of the 12 features, and they are not used in the development of the HDL predictor. In the main text, our numerical results indicate that predicting disruptions across different subdivisions of HP data (i.e. low $B_{tor}$ and high $B_{tor}$) can be difficult. The PCA analyses here further support these results. As shown in the below figures, the discrepancy in $B_{tor}$ ranges (see table 2 in the main text and table 1-2 in appendix for the ranges of $B_{tor}$) produces a large separation between the resulting HP low $B_{tor}$ and HP high $B_{tor}$ plasmas (A8(a), A9(a), A10(a)) which can make the predictor trained with HP low $B_{tor}$ data work poorly on HP high $B_{tor}$ data. Furthermore, if we compare subplots a, b of A8-A10, removing the $P_{in}$ constraint in HP low $B_{tor}$ data (i.e. high $\beta_p$, low $q_{95}$, low $B_{tor}$ data) slightly increases its overlap with HP high $B_{tor}$ data. And indeed, our results suggest that a predictor trained with high $β_p$, low $q_{95}$, low $B_{tor}$ data achieves better accuracy on HP high $B_{tor}$ plasmas (case 4-5 in figure 2, case 4-5 in appendix figure 4, 6). The results from these PCA plots also confirm the tight correlation among signals related to disruption prediction. Although $B_{tor}$ and $P_{in}$ are not directly involved in PCA transformation, limiting their ranges does greatly change the distribution of 12 training features. Thus, constraining the ranges of a few parameters related to disruption prediction can give clear separation between the resulting different subdivisions of the plasmas, whether these parameters are directly used in the development of predictor or not.

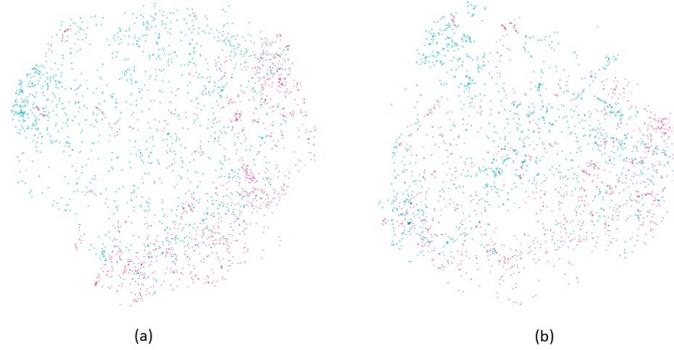

**Figure 8: The PCA clustering plots for different subdivisions of C-Mod HP data: (a) HP low $B_{tor}$ (cyan) vs. HP high $B_{tor}$ (magenta); (b) high $β_p$, low $q_{95}$, low $B_{tor}$ (cyan) vs. HP high $B_{tor}$ (magenta). The coloring is done *a posteriori*.**

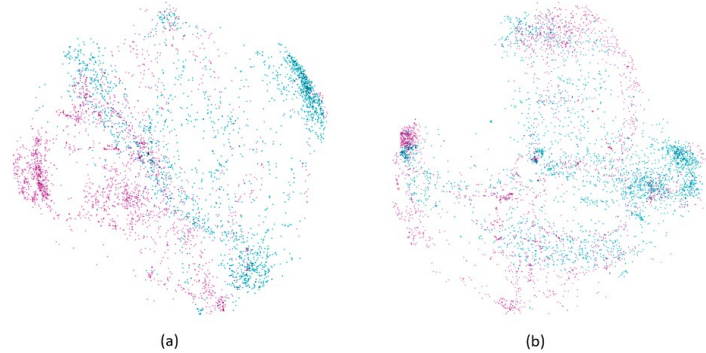

**Figure 9: The PCA clustering plots for different subdivisions of DIII-D HP data: (a) HP low $B_{tor}$ (cyan) vs. HP high $B_{tor}$ (magenta); (b) high $β_p$, low $q_{95}$, low $B_{tor}$ (cyan) vs. HP high $B_{tor}$ (magenta). The coloring is done *a posteriori*.**

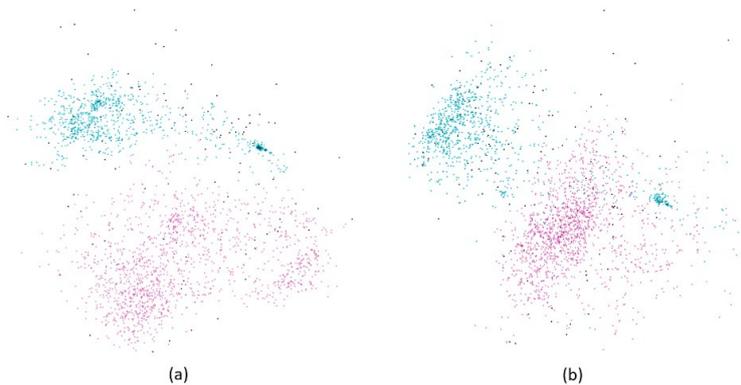

**Figure 10: The PCA clustering plots for different subdivisions of EAST HP data: (a) HP low $B_{tor}$ (cyan) vs. HP high $B_{tor}$ (magenta); (b) high $β_p$, low $q_{95}$, low $B_{tor}$ (cyan) vs. HP high $B_{tor}$ (magenta). The coloring is done *a posteriori*.**